\documentclass[12pt,preprint]{aastex}
\usepackage{emulateapj5}

\newcommand{\lopt}{\ifmmode L_{2500} \else $~L_{2500}$\fi}
\newcommand{\loglopt}{\ifmmode{\rm log}~L_{2500} \else log$~L_{2500}$\fi}
\newcommand{\logz}{\ifmmode{\rm log}~z \else log$~z$\fi}
\newcommand{\ew}{\ifmmode{W_{\lambda}} \else $W_{\lamzpcbda}$\fi}
\newcommand{\ax}{\ifmmode{\alpha_x} \else $\alpha_x$\fi} 
\newcommand{\aox}{\ifmmode{\alpha_{\rm ox}} \else $\alpha_{\rm ox}$\fi} 
\newcommand{\fcgs}{\ifmmode {\rm erg~cm}^{-2}~{\rm s}^{-1}\else erg~cm$^{-2}$~s$^{-1}$\fi}
\newcommand{\mfcgs}{\ifmmode {\rm erg~cm}^{-2}~{\rm s}^{-1}{\rm Hz}^{-1}\else erg~cm$^{-2}$~s$^{-1}$~Hz$^{-1}$\fi}
\newcommand{\lcgs}{\ifmmode {\rm erg}~{\rm s}^{-1}\else erg~s$^{-1}$\fi}
\newcommand{\mlcgs}{\ifmmode {\rm erg}~{\rm s}^{-1}{\rm Hz}^{-1}\else erg~s$^{-1}$\fi}
\newcommand{\kms}{\ifmmode~{\rm km~s}^{-1}\else ~km~s$^{-1}~$\fi}
\newcommand{\mone}{\ifmmode ^{-1}\else$^{-1}$\fi}
\newcommand{\mtwo}{\ifmmode ^{-2}\else$^{-2}$\fi}
\newcommand{\lapprox }{{\lower0.8ex\hbox{$\buildrel <\over\sim$}}}
\newcommand{\gapprox }{{\lower0.8ex\hbox{$\buildrel >\over\sim$}}}
\newcommand{\msun}{\ifmmode {M_{\odot}}\else${M_{\odot}}$\fi}
\newcommand{\nh}{\ifmmode{\rm N_{H}} \else N$_{H}$\fi}
\newcommand{\nhgal}{\ifmmode{ N_{H}^{Gal}} \else N$_{H}^{Gal}$\fi}
\newcommand{\nhintr}{\ifmmode{ N_{H}^{intr}} \else N$_{H}^{intr}$\fi}
\newcommand{\nhtot}{\ifmmode{ N_{H}^{tot}} \else N$_{H}^{tot}$\fi}
\newcommand{\atoms}{\ifmmode{\rm ~atoms~cm^{-2}} \else ~atoms cm$^{-2}$\fi}
\newcommand{\cmsq}{\ifmmode{\rm ~cm^{-2}} \else cm$^{-2}$\fi}
\received{}
\accepted{}

\lefthead{Green et al.}
\righthead{Q2345+007: Binary Quasar or Massive Dark Lens?} 

\slugcomment{ApJ, accepted Feb 1, 2002}
\shorttitle{QSO Pair Q2345+007}
\shortauthors{Green et al.}

\begin{document}
\title{Chandra Observations of the QSO Pair Q2345+007: \\
Binary Quasar or Massive Dark Lens?} 

\author{Paul J. Green\altaffilmark{1,2},
Chris Kochanek, 
Aneta Siemiginowska, 
Dong-Woo Kim, 
Maxim Markevitch, 
John Silverman\altaffilmark{1},
Anil Dosaj\altaffilmark{1}, } 
\affil{Harvard-Smithsonian Center for Astrophysics, 60 Garden St.,
Cambridge, MA 02138} 

\author{Buell T. Jannuzi\altaffilmark{1}} 
\affil{National Optical Astronomy Observatory, P.O. Box 26732,
Tucson, AZ 85726-6732} 

\author{Chris Smith\altaffilmark{1}} 
\affil{Cerro Tololo Inter-American Observatory, National Optical
Astronomy Observatory, Casilla 603, La Serena, Chile.} 

\altaffiltext{1}{Visiting Astronomer, Kitt Peak National Observatory
and/or Cerro Tololo Inter-American Observatory, National
Optical Astronomy Observatory, operated by the Association
of Universities for Research in Astronomy, Inc. (AURA) under
cooperative agreement with the National Science Foundation.}
\altaffiltext{2}{email: pgreen@cfa.harvard.edu}

\begin{abstract}
The components of the wide ($7\arcsec$.3) separation quasar pair
Q2345+007A,B ($z=2.15$) have the most strikingly similar optical
spectra seen to date (Steidel \& Sargent 1991) yet no detected lensing
mass, making this system the best candidate known for a massive
($\sim10^{14}M_\odot$) dark matter lens system. Here we present results
from a 65\,ksec {\em Chandra} observation designed to investigate
whether it is a binary quasar or a gravitational lens.  We find no
X-ray evidence for a lensing cluster to a (0.5-2keV) flux limit of
$2\times 10^{-15}\fcgs$, which is consistent with lensing only for a
reduced baryon fraction. Using the Chandra X-ray observations of the
quasars themselves, together with new and published optical
measurements, we use the observed emission properties of the quasars
for further tests between the lens and binary hypotheses.  Assuming
similar line-of-sight absorption to the images, we find that their
X-ray continuum slopes are inconsistent
($\Gamma_A=2.30^{+0.36}_{-0.30}$ and $\Gamma_B=0.83^{+0.49}_{-0.44}$)
as are their X-ray to optical flux ratios. The probability that B
suffers absorption sufficient to account for these spectral
differences is negligible.  We present new optical evidence that the
flux ratio of the pair is variable, so the time-delay in a lens
scenario could cause some of the discrepancies.  However,
adequately large variations in overall spectral energy distribution
are rare in individual QSOs.  All new evidence here weighs strongly
toward the binary interpretation.  Q2345+007 thus may represent the
highest redshift example known of interaction-triggered but as-yet
unmerged luminous AGN.
\end{abstract}

\vskip1cm
{\em Subject headings:} gravitational lensing -- quasars: absorption
lines -- quasars: individual (Q2345+007) -- X-rays: general -- X-rays:
individual (Q2345+007) 


\section{Introduction}
\label{intro}

The density and evolution of massive halos ($M >10^{14}M_\odot$) is a
powerful probe of the cosmological model.  The number of massive
clusters depends exponentially on the amplitude of the power spectrum
when normalized by $\sigma_8$, the amplitude on a scale of
$8h^{-1}$~Mpc.  These halos can be detected through an overdensity of
galaxies (e.g. Postman et al. 1996 for a modern example), from X-ray
emission (e.g. Ebeling et al. 2000), or with the Sunyaev-Zeldovich
effect (e.g. Holder et al. 2000).  All these methods depend on
emission or absorption from the baryons in the halo.  Alternatively,
we can detect such massive halos using gravitational lensing.  This
can be done either with weak lensing surveys (Wittman et al. 2001) or
surveys for multiply imaged background sources with wide image
separations (usually defined by $\Delta\theta > 3\arcsec$;
Kochanek, Falco \& Munoz 1999, hereafter KFM99).

Unlike the other methods of finding clusters, gravitational lensing
can also find ``dark'' clusters where there is mass but no 
detectable baryons.  Some candidates have been found in weak lensing
(shear-selected) surveys (Umetsu \& Futamase 2000; Erben et al. 2000), but
only one such  cluster has been spectroscopically confirmed to date
(Wittman et al. 2001).  Most of the debate about dark halos has
focused on the population of wide-separation quasar pairs (WSQPs).
All the lens candidates with separations $\Delta\theta < 3\arcsec$
have identifiable primary lens galaxies in deep NICMOS
observations (see synopses in KFM99, or Mortlock \& Webster
2000). Above $3\arcsec$ there are 17 objects.  Four are {\em bona
fide} gravitational lenses, with similar optical/radio flux ratios and
negligible spectroscopic differences, as well as a plausible lensing
galaxy.  Eight more pairs, with discrepant radio/optical flux ratios
for a lens, or $>3\sigma$ 
velocity differences, are very probably binary quasars. The remaining 
5 objects are WSQPs with similar spectra, essentially identical
redshifts, and no visible lens galaxy or cluster. Were four 
pairs found to be dark clusters, they would imply that the dark
clusters were just as numerous as normal clusters.      

The problem with simply interpreting the WSQPs as dark clusters is
that we expect to find WSQPs even in the absence of dark lensing as
examples of binary quasars in which the central engines of two nearby
galaxies are simultaneously active.  We know that many of the pairs
are binary quasars because they have discrepant flux ratios as a
function of wavelength or $>3\sigma$ velocity differences.  We can
prove from statistical analyses of the relative numbers of
radio-quiet/radio-quiet, radio-quiet/radio-loud and 
radio-loud/radio-loud pairs that most of the remaining WSQPs are
binary quasars (KFM99).  Since the number of WSQPs is
100 times higher than expected from simple extrapolations of the
quasar-quasar correlation function, galaxy interactions are crucial to
creating binary quasars (see KFM99, Mortlock \& Webster
2000) and can be used as a tool to study the triggering of nuclear
activity in galaxies (e.g. Osterbrock 1993).  

Nonetheless, we still have puzzling examples of WSQPs whose members have 
startlingly similar optical spectroscopic properties but no evidence
for a lens. One example is the optical quasar pair Q2345$+$007A,B
(Tyson et al. 1986).  With a separation of 7$\arcsec$.3, Q2345$+$007
is the most prominent `dark matter'  gravitational lens candidate.
Optical spectra of the two image components show exceptional intrinsic
similarity in 
both line profile and velocity (e.g., $z=2.15$, $\Delta v_{B-A}=15\pm
20$\kms; Steidel \& Sargent 1991; hereafter SS91).  Slight differences
may be explainable as the combined effect of time variability in the
source and time delays produced by the lens (Small et al. 1997).  After
nearly 20 years of study including deep optical and infrared imaging
(Nieto et al. 1988, Weir \& Djorgovski 1991,  Bonnet et al. 1993,
Gopal-Krishna et al. 1993, McLeod et al. 1994, Fischer et al. 1994,
Pello et al. 1996) and VLA radio imaging (Patnaik et al. 1996), this
pair remains the most obstinate mystery.  Either massive, concentrated
dark matter near the line of sight is acting as a gravitational lens,
or two neighboring quasars have virtually identical spectral
characteristics yet significantly different luminosities. 

In this paper we present a deep X-ray image of Q2345$+$007 obtained
with the Chandra X-ray Observatory (CXO).  Using the superb
sensitivity and resolution of the CXO we address two questions.
First, we search the field for extended X-ray emission from hot
baryons in the lens, which allows us to detect an optically dark halo.
We cannot detect a baryon-free halo, since the X-ray emission depends
on the hot baryons in the halo, but we can detect halos with reduced
baryon fractions. Second, we test whether the X-ray flux ratios and
spectra of the images are consistent with the lens hypothesis.
We describe the CXO observations next in \S~\ref{cxo}, and supporting
Kitt Peak optical imaging in \S~\ref{mosaic}.  We 
discuss the search for extended emission from a lensing cluster in
\S~\ref{findlens}, and contrast the properties of the two quasars  in
\S~\ref{abdiffs}, summarizing in \S~\ref{summary}.    

\section{Chandra Observations and Analysis}
\label{cxo}

Q2345+007 was first observed by the CXO on May 26, 2000 for 25.2ksec using
the back-illuminated S3 chip of the Advanced CCD Imaging Spectrometer
(ACIS) in faint, timed-event mode.  We offset the
pointing by 1$\arcmin$ in spacecraft $Y$-coordinate from the Chandra
aimpoint, to allow for imaging of any extended emission all within one
ACIS node.  It was observed again on June 27, 2000 in the same
configuration, for 52.6ksec.  However, because of an incorrectly
configured bias-only run in the previous ACIS segment, event rates
were about 20 times as high as expected from the chips (I1 and I3)
processed by Front End Processors (FEP) 0 and 3.  Due to the telemetry
saturation, this created up to 50\% deadtime in these chips. Telemetry
dropouts are fully accounted for in the final exposure times, but 
the aimpoint chip S3 was in any case not affected. 

We have used data reprocessed (in April 2001) at CXC\footnote{CXCDS
versions R4CU5UPD14.1, along with ACIS calibration data from the
Chandra CALDB~2.0b.}. We then ran XPIPE (Kim et al. 2000) which was
specifically designed for the Chandra Multi-wavelength Project (ChaMP;
Green et al. 2000). Data
screening applied in the CXC Level 2 processing excludes events with
bad grades (mostly cosmic ray events) and events with status bits set
such as bad pixels and columns. Additional bad pixels and columns were
excluded by examining events in the chip coordinates in each chip. To
remove time intervals of high background rates, we make a light curve
and exclude those time intervals with 3-$\sigma$ or greater
fluctuation above the 
mean background count rate.  Given different characteristics between
BI and FI chips, this is done separately for each ACIS chip.  The
remaining exposure time for the back-illuminated S3 chip is 64,998 sec.

For source detection, we have applied the {\tt wavdetect} algorithm
(Freeman et al. 2002), available in the {\em Chandra Interactive Analysis
of Observations} (CIAO) software package.\footnote{CIAO may be
downloaded from {\tt http://asc.harvard.edu/ciao/.}}  {\tt Wavdetect}
is more reliable than the traditional {\tt celldetect} algorithm
for finding individual sources in crowded fields and for identifying
extended sources, but the algorithm tends to detect spurious
sources near the detector edge.  To avoid such false detections, we 
generate an exposure map for each chip (assuming a monoenergetic
distribution of source photons at 1.5keV) and apply an exposure
threshold parameter to be 10\%.  After performing various tests to
find the most efficient parameters (see Kim et al. 2000), we select a
significance threshold parameter to be $10^{-6}$, corresponding to 1
spurious source per CCD chip.  We use scale parameters of 1, 2,
4, 8, 16, 32, and 64 pixels to cover a wide range of source scales,
thus accommodating the PSF variation as a function of off-axis
angle. For other parameters, we have used default values given in
CIAO2.1. 

\begin{figure*}[ht]
\figurenum{1}
\centering
\vspace*{0.0in}
\plotone{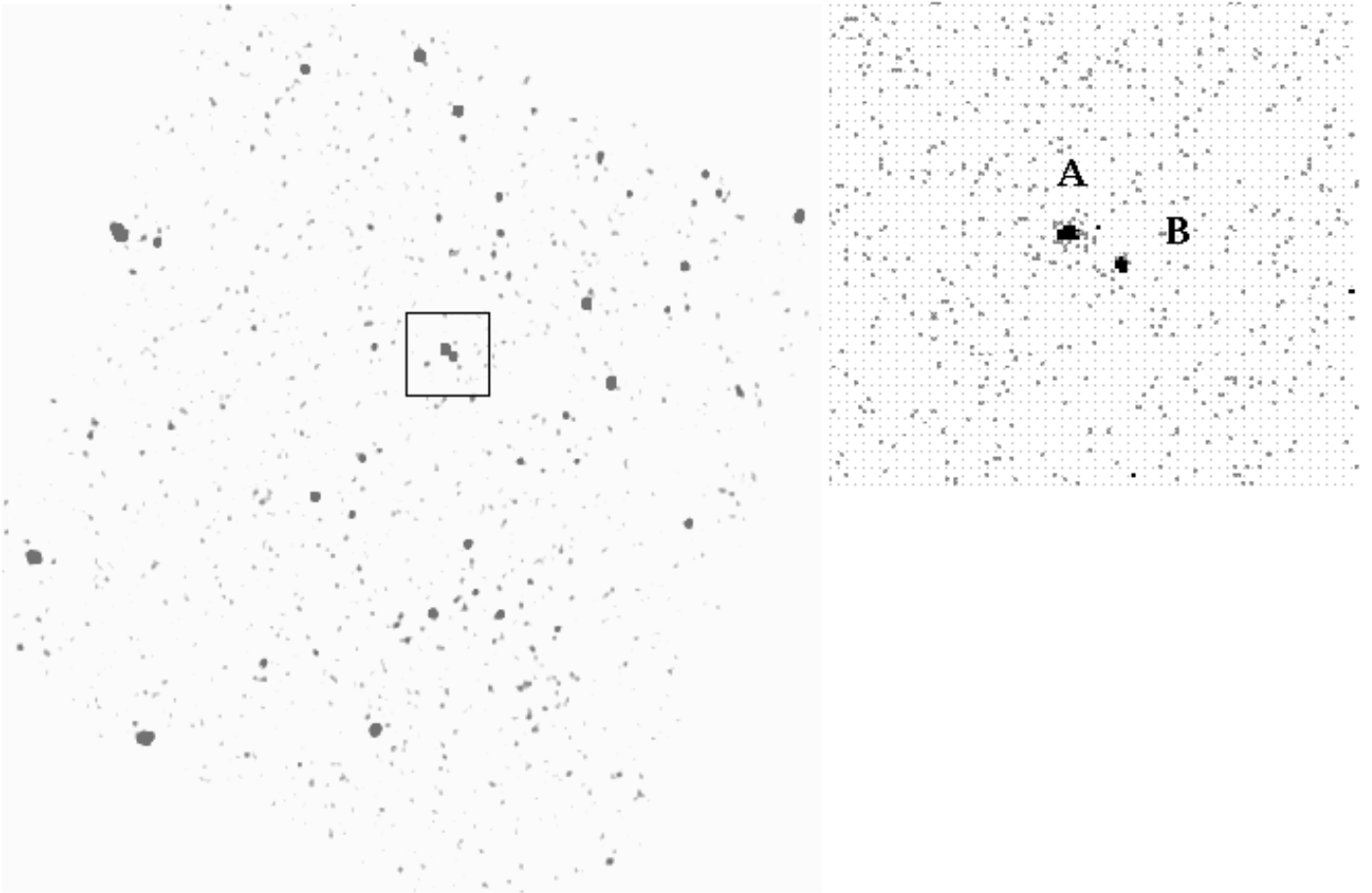}
\caption{\small 
LEFT: This $9\arcmin \times 9\arcmin$ ACIS-S3 (0.3-8keV) image of
Q2345+007 (North up, East to the left) was smoothed with a 3 pixel
(1.5$\arcsec$) gaussian for clarity, and shows the QSO pair on chip 
S3, about 1$\arcmin$ North and 30$\arcsec$ West of the chip center.  While
numerous point sources are detected, no extended cluster emission is
distinguishable around the quasar pair on the image.  RIGHT: A
$1\arcmin \times 1\arcmin$ close-up of Q2345+007A,B shows that the
pair is well-resolved, with no detectable emission from nearby point
sources.    
\label{fs3}}
\end{figure*}

As shown in Figure~\ref{fs3}, Q2345+007 A and B are well-resolved and
strongly detected in the image.  The Chandra astrometry for the
pair corresponds closely ($\sim0.3\arcsec$) to the optical counterpart
positions, as expected from Chandra's absolute aspect quality
(Aldcroft et al. 2000).  We perform our own aperture photometry at the
{\tt wavdetect} positions of the 46 detected sources on this chip,
which yields sources between 7 to 608 net counts (0.5-8\,keV), and S/N
from 1.6 to 24.     

\section{MOSAIC Optical Imaging Observations}
\label{mosaic}

Several lensing cluster candidates have been suggested from analysis
of deep optical images. As part of the Chandra Multiwavelength Project
(ChaMP) for followup of Chandra serendipitous sources (Green et
al. 2000), on UT 29 September, 2000 and 22 Aug 2001, we obtained
images in Sloan $g^\prime$, $r^\prime$, and $i^\prime$ filters at the
CTIO 4meter Blanco telescope using the wide-field MOSAIC camera, which
has eight 2048$\times$4096 chips in a 4$\times$2 array comprising a
36$\times$36\arcmin\, field of view.  We reduced the images using the
MSCRED (v4.1) package (Valdes \& Tody 1998; Valdes 1998) in the IRAF
environment (Tody et al. 1986).  This proceeds via the usual initial
calibration, subtracting median-combined bias frames and dividing by
similarly prepared dome flats.  Additionally, we correct for
electronic crosstalk between pairs of CCDs sharing readout
electronics.  We then make a single image in each filter by
median-combining multiple object frames, effectively rejecting all of
the celestial objects in the final frame.  From this we create a
fringe-correction frame in each filter with the large-scale variation
removed, and interactively subtract scaled versions from every
exposure in each given filter.  We then generate the sky flat for
large scale corrections, again via median filtering, this time of
fringe-corrected object frames.  This sky flat is divided into every
object frame. 

From our MOSAIC images, we determined the (previously unpublished)
positions of optically-identified lens candidates discussed below,
listed for convenience in Table~1. These 
optical positions are based on an astrometric solution of rms
$\sim0.1\arcsec$ obtained by matching detected objects to the GSC2.2
catalog.\footnote{The Guide Star Catalogue-II is a joint project of the
Space Telescope Science Institute and the Osservatorio Astronomico di
Torino. Space Telescope Science Institute is operated by the
Association of Universities for Research in Astronomy, for the
National Aeronautics and Space Administration under contract
NAS5-26555.} 
On each observing run, we obtained for this field 3 dithered images
per filter.  After we flag bad pixels and charge bleeds from severely
saturated stars in each image, we project them onto the tangent-plane.
Removal of remaining large-scale gradients 
and scale differences 
between dithered images allows for combination into a final single
stacked image in each filter 
These nights were probably photometric, since the scaling values were
close to one.  

We used the MOSAIC images to determine the optical flux ratios
of the quasar pair in our 3 filters.  These are tabulated
as magnitude differences in Table~2, along with previously
published values, for convenience.  We discuss the evident variability
in \S~\ref{aox}.

\section{The Search for the Lens}
\label{findlens}

We now search for X-rays from any extended halo centered
on the WSQP, and also measure X-rays from published optical cluster
candidates.   In \S~\ref{extent} we describe our search for X-ray
emission from any lensing halo, and in \S~\ref{clusters} we compare
the extended X-ray emission to the optically-identified cluster
candidates inferred from the distribution of galaxies in the field
(Bonnet et al. 1993; Pello et al. 1996).  In \S~\ref{lenscluster} we 
discuss the implications of our measurements for the lens hypothesis. 

\subsection{Search for Extended X-ray Emission }
\label{extent}

\begin{figure*}[ht]
\figurenum{2}
\plotone{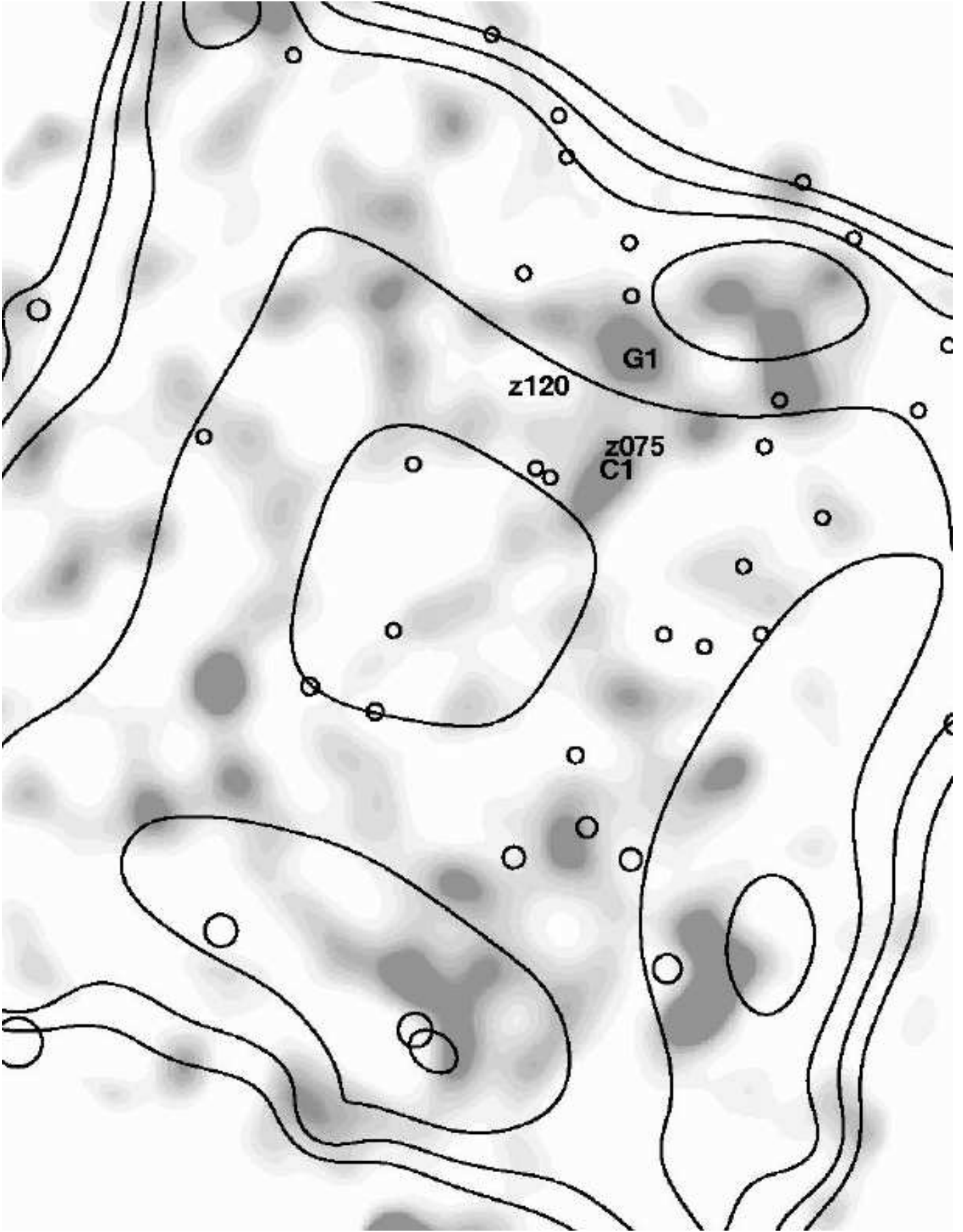}
\caption{Chandra ACIS-S3 Image of the $9.7\arcmin$ field surrounding
Q2345$+$007, with North up, East to the left. We removed counts within
these apertures, and then divided by an exposure map before smoothing
with a 10$\arcsec$ gaussian.  The resulting greyscale shows a mean
flux of $\sim2\times~10^{-10}~photons~cm\mtwo~sec\mone~pixel\mone$,
with values across the image ranging from about 0.7 to 5.3 in those
units. Circles with sizes representing the PSF (95\% encircled energy)
mark the positions of point sources detected by CIAO {\tt wavdetect}.  The
positions of the twin QSOs are evident just NE of center. The large
contours show linear levels (from 500 to 900cm$^2$, in steps of 100)
in the exposure map. Positions of putative optically-identified galaxy
clusters are  marked in bold type.   
\label{fnosrc} }
\end{figure*}

No significant extended emission sources are evident to the eye on the
ACIS-S3 image.  When searching for faint extended sources, 
however, it is important to minimize background contamination.
The ACIS particle background increases significantly below 0.5keV
and again at high energies.  To optimize detection of a weak cluster
signal, we first filtered the cleaned, combined image
to include only photons between 0.5 and 2keV.  We then masked out
pixels within a radius encompassing 95\% of the encircled energy
around all point sources detected by {\tt wavdetect}.  This image is
divided by the appropriate 
ACIS-S3 exposure map, which takes account instrumental features
(effective area, quantum efficiency, telescope vignetting) as well as
relative exposure due to the dither pattern.  Further division by the
exposure time yields a normalized image with pixel values in
photons~cm$^{-2}$~s$^{-1}$.  To facilitate the search for nearby
extended sources, we smoothed the point-source-subtracted image using
a 10$\arcsec$ FWHM Gaussian.   Figure~\ref{fnosrc} shows several
marginal excesses at $\sim20\%$ above the background level of
$2\times 10^{-10}$photons~cm\mtwo~sec\mone~\arcsec\mtwo
(corresponding to 0.16 counts/sec across the entire chip).  Several
low-level ($\sim 10\%$) features in the normalized image in
Figure~\ref{fnosrc} partially coincide with features in the exposure
map.  This may be because the telescope effective area varies in the
0.5-2keV bandpass and the assumed spectrum used to compute the
exposure map (monoenergetic 1.5keV)  may not represent the incident
spectra very well across the image.   More detailed simulations
for detecting extended structures in ACIS images are called for, but
beyond the scope of this paper. None of the features were detected by
our wavelet detection analysis, so we do not consider any of the
apparent fluctuations to be significant unless they appear clearly in
radial profiles.

In Table~1, we tabulate the X-ray positions (based on
the peak flux in the smoothed image) of apparent extended excesses near
Q2345+007.  We summed the X-ray counts in 10$\arcsec$ annuli centered on
a point midway between the two QSO images, and extending to a
2$\arcmin$ radius, and detect no excess over the background. To derive
an upper limit on the extended source flux, we denote source
counts $S$ and background counts $B$ (where $B$ has been 
estimated from an area $A_B$ but normalized to the source area $A_S$).
\footnote{Source counts are derived from total counts $T$ in area $A_S$ via
$S=T-B$.}  The random $1\sigma$ deviation in counts is
$\sigma= \sqrt{S + kB}$, where $k=(1+\frac{A_S}{A_B})$.
If we define a source count upper limit at $N$ standard deviations,
then our source count upper limit becomes
	$$S= N\sigma = \frac{N}{2} \Bigl(N + \sqrt{N^2 + 4kB}\Bigr)$$
\noindent 
We estimate the background from an area $A_B \gg A_S$, and thereby
derive a 95\% upper limit of fewer than 44 counts from any cluster
within a circle of 1$\arcmin$ radius centered on the WSQP,
corresponding to a  
count rate of $7\times 10^{-4}$sec$^{-1}$.   
Considering a Raymond-Smith model with a reasonable range\footnote{The
unabsorbed flux is derived, i.e., from outside the 
Galaxy.  Changes in the assumed spectral models from 2 to 8keV result
in changes of $\sim15\%$ in the value of the derived flux.} of
parameters and Galactic absorption, the   
corresponding upper limit on the $0.5-2$ keV flux is around $2\times
10^{-15}\fcgs.$  This illustrates Chandra's excellent sensitivity to
faint sources, even when extended.\footnote{If there is any residual
flux from the WSQP scattered beyond the radius containing 95\% of the
total energy, our  estimated cluster flux upper limit is conservative.}
 
\subsection{Observed Cluster Constraints}
\label{clusters}


Our 0.5-2keV flux upper limit of $2\times 10^{-15}\fcgs$ within 1\arcmin\,
of the WSQP clearly yields a stringent limit on the existence of any lensing
cluster. Assuming a $T=2-3$ keV spectrum, this upper limit corresponds at
$z\sim1$ to a $0.5-2$ keV (rest frame) luminosity of $1.2\times
10^{43}\lcgs$ ($1.6\times
10^{43}$ for $\Omega=0.3$). An $r=1'$ aperture at $z=1$
corresponds to 0.5Mpc (0.6Mpc for $\Omega=0.3$) and encloses
about a half of the total luminosity for a typical cluster with the surface
brightness described by a $\beta$-model with a core radius of 0.25 Mpc and
$\beta=0.6$ (Jones \& Forman 1984). With this correction, 
our upper limit corresponds to a typical luminosity of 
a galaxy group with $T\sim 2-3$ keV (e.g., Hwang et al. 1999 and
references therein), so our estimate is self-consistent.
All luminosities, sizes and distances are calculated assuming
$H_0=50$\,km\,s\mone~Mpc\mone, $\Omega_0=1.0$, and $\Lambda=0.0$
unless otherwise noted (we also give $\Omega_0=0.3$ values).  This
represents by far the strongest constraint on the X-ray luminosity of
any dark lens candidate to date (cf., Chartas et al. 2001). 

It is certainly possible that a massive cluster that is {\em not}
centered on the WSQP could also produce the observed image splitting. 
Therefore, we also investigate apparent flux excesses in the vicinity
($<1.5\arcmin$) of the quasar pair.  However, the required mass of the
lensing cluster would rise with angular distance from the WSQP, so that our
flux limits constrain the existence of such a cluster even more
strongly at larger angles.

\begin{figure*}[ht]
\figurenum{3}
\plotone{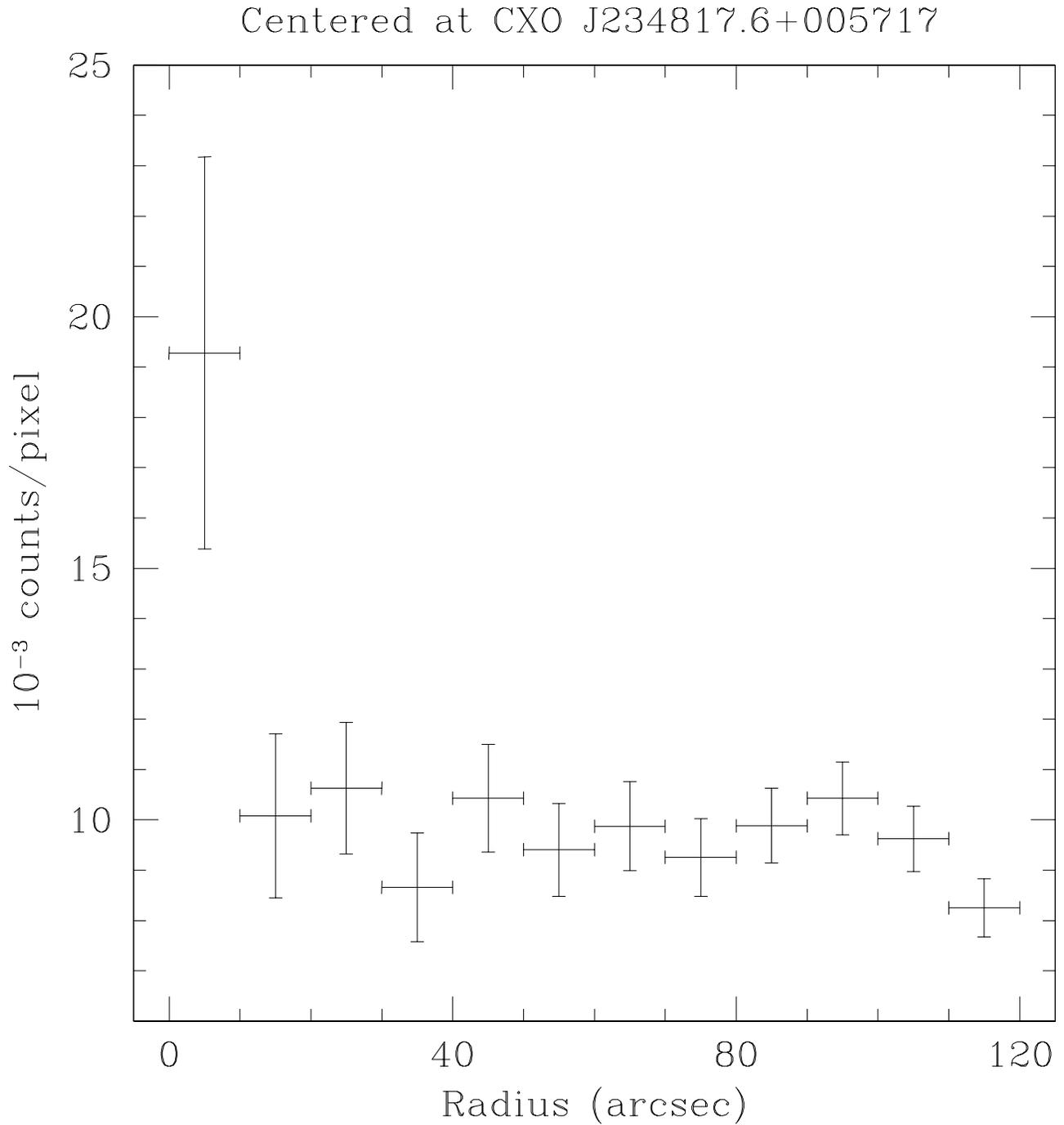}
\caption{This radial profile is centered at the position of 
CXO J234817.6+005717, the excess nearest the QSOs in
Figure~\ref{fnosrcopt}.  We used successive 10$\arcsec$ annuli
in a cleaned 0.5-2kev image, excluding both detected point sources and
regions with apparent source excesses in Figure~\ref{fnosrc}. Vertical
error bars are 1$\sigma$, and horizontal bars represent the range of
each annulus. The background level ($\sim0.01$ counts per $0.5\arcsec$
pixel) has not been subtracted. The profile displays a possible flux
excess in the first bin, significant at 2.5-3$\sigma$ above background.
\label{fprof} }
\end{figure*}

The closest significant excess of extended X-ray emission to the WSQP
is CXOJ234817.6+005717.  Centered about 27$\arcsec$ from the center of
the WSQP, it has a peak flux of about 0.0544 counts cm\mtwo s\mone
arcsec\mtwo.  
The radial profile of this excess, determined with possible nearby
extended sources removed, is shown in Figure~\ref{fprof}. The excess
contributes just $11\pm5$ counts above background.  At just
2.2$\sigma$, we find this source to be of questionable certitude.  A
more credible $2\sigma$ upper limit to the flux yields $9\times
10^{-16}\fcgs$ from 0.5-2keV. 

Bonnet et al. (1993) identified a cluster center they label {\bf C1}
from a weak gravitational shear field pattern suggesting a lens
velocity dispersion $\sigma_v\sim1200$\kms. The peak X-ray flux of the
above extended source corresponds to a position just 8$\arcsec$ from
C1.  Pello et al. (1996) claimed an excess of galaxies with
photometric redshifts $z\sim0.75$, which is 13$\arcsec$ from C1 and
21$\arcsec$ from CXOJ234817.6+005717.  However, centroids may
naturally have somewhat different positions because optical galaxies
may not follow the overall mass distribution, and the cluster also may
not be virialized.  Assuming that all 3 objects
(C1, the Pello et al. z=0.75 optical galaxy excess, and
CXOJ234817.6+005717) can be identified as the same object, 
the $0.5-2$ keV luminosity upper limit is
$2.8\times 10^{42}\lcgs$ ($3.6\times 10^{42}\lcgs$ for $\Omega=0.3$)
more similar to an isolated elliptical galaxy or a small group than to a
rich $\sigma_v \sim 1200$\kms\ cluster, suggesting that the large velocity
dispersion estimate is the result of a line of sight projection.
In any case, at a transverse distance of $\sim 220h_{50}$kpc, such a 
mass is too small to produce the observed pair separation by
lensing.  

Bonnet et al. also identified the galaxy G2\footnote{G1 is the
brighter galaxy.  Labels in Bonnet et al. (1993) for G1 and G2 are
incorrectly swapped in all but their Figure~2.} as having a position
consistent with C1 within the errors ($\sim9\arcsec$ from the best
centroid of C1).  At 15$\arcsec$ from G2, CXOJ234817.6+005717 is
clearly not consistent with emission from that galaxy. 
Another apparent X-ray excess CXOJ234816.9+005811 is an arcminute
from the WSQP centroid, but 7$\arcsec$ from the galaxy G1 identified
by Bonnet et al. (1993).  However, our radial profile centered either on
CXOJ234816.9+005811 or on the position of G1 reveals no significant
excess above background, even when the other nearby extended sources
are excised.   

\subsection{Comparison to Required Lens}
\label{lenscluster}

We start by assuming the WSQP is a lens produced by a simple, singular
isothermal sphere (SIS, see Schneider, Ehlers \& Falco 1992).  
The image separation $\Delta\theta=8\pi(\sigma_v/c)^2 D_{LS}/D_{OS}$ 
depends only on the velocity dispersion of the potential $\sigma_v$
and the ratio of the comoving distances between the 
lens and the source, $D_{LS}$, and the observer and the source,
$D_{OS}$.  Since we failed to detect a lens cluster, we will get the most 
conservative limits if we assume the lens lies at the ``minimum
flux redshift,'' the redshift that would minimize the observed
X-ray flux. If we neglect K-corrections, the flux from the lens is  
\begin{equation} 
  F =  { L \over 4\pi D_{OL}^2(1+z_l)^2 }  \propto
    \left[ { D_{OS} r_H \over D_{OL} D_{LS} (1+z_l) } \right]^2 
\end{equation}
\noindent 
where $r_H$ is the Hubble radius $c/H_0$. 
This flux diverges at low redshift because of the proximity of the cluster
and at high redshift because of the mass of the cluster.  For $\Omega_0=1$
the flux is minimized at $z_l=0.92$, which we will round to $z_l=1$
for simplicity.

Now using a SIS model for the lensing mass, the large image separation
of  7$\arcsec$.3 implies a cluster velocity dispersion of
$\sigma_v=860 \kms$ or a cluster mass of 
$1.3\times 10^{14}\msun$. We emphasize that this is the {\em minimum}
enclosed mass or dispersion required at this redshift to induce the
observed pair separation. Combining the $L_X-\sigma_v$ relation from,
e.g., Mulchaey \& Zabludoff (1998) and the $L_X-T$ relation from,
e.g., Markevitch (1998), and neglecting any possible cosmological
evolution of these relations for a qualitative estimate, we obtain
$L_X (0.5-2\,{\rm keV})\approx 2\times 10^{44}$\lcgs\, and $T\approx
5$ keV for such a cluster. At $z=1$, this corresponds to $f_X
(0.5-2\,{\rm keV})\approx 4\times 10^{-14}$\fcgs\, ($3\times 10^{-14}$
for $\Omega=0.3$). Again assuming a typical cluster brightness
distribution and dividing by 2 to convert to the $r=1'$ aperture, we
can see that {\em our 95\% flux limit is an order of magnitude below this
minimum required flux estimate}. If the lens were a dark cluster lacking not
only galaxies but also gas, the gas fraction would have to be $\sim 3$
times lower than that in known clusters. At low redshift, all well-studied
clusters (at least in the relevant range of radii and masses) have
similar values of the gas fraction (e.g., Mohr, Mathiesen, \& Evrard
1999; Vikhlinin, Forman, \& Jones 1999) so such a deviation appears
to be extremely unlikely. We conclude that a single cluster acting as
a lens is not a plausible scenario for Q2345+007.

In other lenses clearly due to a combination of a cluster and a galaxy
(particularly Q0957+561; Keeton et al. 2000), a massive, luminous lens
galaxy dominates the image splitting.  Here we see no such candidate
galaxy, even in the infrared, to a limit of approximately $L_*/10$
near redshift unity. Such a galaxy, unless completely different from
all other known lens galaxies (e.g., Kochanek et al. 2000,
Xanthopoulos et al. 1998), must make a negligible contribution to the
overall image separation and  modifies our estimate of the critical
radius of the putative dark halo only by a factor of
$1-2b_{gal}/\Delta\theta$ with $b_{gal}\ll 1\farcs0$.  


\section{Are They Images of the Same Quasar?}
\label{abdiffs}

We have failed to find a lens. While a dark matter lens
is not ruled out, the pair could also be shown to be a
binary quasar by evidence that the components' spectral energy
distributions are different.  We explore this by first determining the
X-ray properties of the two quasars in \S~\ref{qsox}, followed by a
discussion of the possible effects of aborption and extinction on the
flux ratios in 
\S~\ref{qsoabs}.  In \S~\ref{aox} we compare the flux ratios from the
near infrared to the X-ray bands to conclude that they are probably
different quasars.  We discuss the remaining puzzle of the strikingly
similar optical/UV spectra of the components in \S~\ref{ospec}.

\subsection{The X-ray Properties of the Quasars}
\label{qsox}

The total broad band (0.3-8keV) counts from the QSOs are
$358.5\pm20$ and $54.8\pm8$ for components A and B, respectively.
The broad band X-ray flux ratio is $6.55\pm0.50$.  

As a first check of whether the QSOs have similar
X-ray spectra, we measure their hardness ratio, $HR=(H-S)/(H+S)$,
using the hard counts (H) from 2.5 - 8.0keV and the soft counts (S)
from 0.3 - 2.5keV. While differences in instrument and telescope
calibration are completely negligible at these spacings, the benefit
of comparing hardness ratios directly are that no spectral models
need be fitted to the data.  We find $HR_A=-0.89\pm0.02$ and 
$HR_B=-0.55\pm0.11$, which are significantly different (at
$3\sigma$).   

However, although optical magnitudes suggest little differential
reddening (see \S~\ref{qsoabs} below), absorption along different
sightlines to the QSO might still produce a significant difference in
$HR$ if the absorbers are not dusty.  We also calculate the X-ray flux
ratio in a bandpass that is much less subject to absorption.
In the 1.6-8~keV band (5-25~keV in the QSO restframe) the X-ray flux
ratio is smaller, $2.3\pm 0.5$.  This large difference from the
broadband flux ratio is expected given their differing hardness
ratios.  

A test for differential absorption requires X-ray
spectral modeling. For spectral analyses, we used the latest
(CALDB2.7) ACIS-S3 FITS 
Embedded Function (FEF) files and corresponding gain tables. The new
FEFs are analytic fits based on a physical model of the
back-illuminated CCD, developed by the MIT ACIS/IPI team, and adjusted
to the on-orbit gain of the S3 chip as determined by the flight
calibration sources.  These FEFs are used to generate a Response
Matrix File (RMF), which maps the incident photon energy to ACIS pulse
height (deposited charge).  An Ancillary Response File (ARF)
calibrates the effective collecting area of a specified source region
on ACIS as a function of incident photon energy.  We extract ACIS PI
spectra from a 3.5\arcsec\, region around each QSO, using the
\texttt{psextract} script described in the standard thread for
CIAO2.1.  This script creates an aspect histogram file, and the RMF
and ARF calibration files appropriate to the source position on chip
(which is time-dependent due to dither) and CCD temperature
($-120\,$C).

We first group the Chandra spectra into bins containing at least 10
counts each.  We test four spectral models to test permutations
that link or decouple the power-law and absorbing columns of
Q2345+007A and B.  Results for all four models are compiled in
Table~3. Model (1) fixes ($z=0$) absorption at the Galactic 
value, but allows the power-law slope and flux normalization to vary
for each QSO, as 
$$ N(E) = A_i\, E^{-\Gamma_i}\,e^{-N^{Gal}_{H}\sigma(E)}
~{\rm ~photons~cm^{-2}~s^{-1}~keV^{-1}}$$
In this formula, $A_i$ is the normalization and $\Gamma_i$ the {\em
global} power-law photon index for the individual QSOs.  $N_{H}$
is the equivalent Galactic neutral hydrogen column density 
    $3.81\times 10^{20}\cmsq$ 
which characterizes the effective absorption (by cold 
gas at solar abundance), with $\sigma(E)$ the
corresponding absorption cross-section (Morrison \& McCammon 1983).
For determining the best-fit parameter values, we use Powell
optimization with Primini statistics (Kearns et al. 1995).  This fit
yields a typical 
continuum slope $\Gamma_A=2.19\pm 0.15$ for the brighter
component ($\chi_{\nu}^2=1.1$).  The fainter QSO shows a best-fit of  
$\Gamma_B=0.79\pm 0.4$ ($\chi_{\nu}^2=0.6$).  These confirm our
impression from the hardness ratio that the observed energy
distributions are not consistent.

There are not sufficient counts from B to independently fit both its
absorbing column and power-law slope.  However, a somewhat stronger
test than Model (1) is available by assuming that A and B have
identical total absorbing columns, fitting that column simultaneously
(Model 2).  The result is a column $N_H=5.3\pm3.1 \times
10^{20}\cmsq$, which yields slopes $\Gamma_A=2.30^{+0.36}_{-0.30}$ and
$\Gamma_B=0.83^{+0.49}_{-0.44}$. Therefore, {\em if} the X-ray
emission from A and B suffer identical absorption along the line of
sight, we can confidently say that their intrinsic spectral energy
distributions are inconsistent.  Contour plots for this simultaneous
power-law fit are shown in Figure~\ref{fspeccont}.  Not surprisingly
(since A dominates the total counts) we find  similar results when
using for B the  best-fit column from A alone.

Rather than accepting the power-law difference as intrinsic,
we might entertain the possibility that B suffers from strong
absorption, since this can mimic a flat power-law slope in low S/N
X-ray spectra. If we assume that the system is lensed, then the
underlying X-ray  power-law continuum slopes should be
indistinguishable, even though their line-of-sight absorption may
vary.  We thus fit Model (3) by assuming a single power-law slope, but
allowing for different absorbing columns between the two
components. The best-fit continuum slope is
$\Gamma=2.14^{+0.34}_{-0.28}$, and the absorbing columns are
quite different:  
	$N_{H,A}=3.4^{+3.5}_{-2.9} \times 10^{20}\cmsq$ and
	$N_{H,B}=43.9^{+29}_{-18} \times 10^{20}\cmsq.$ 
Contour plots for this simultaneous absorption column fit are shown in
Figure~\ref{fspeccont}.

If we also allow the absorption columns to differ between A and B
(Model 4), the difference between the best-fit power-law slopes
becomes insignificant.  This is because, as expected, the small
number of counts in B are insufficient to constrain both parameters.  
While the parameters for A change little, 
for B, we find $N_{H,B}=23.6^{+28.1}_{-19.2} \times 10^{20}\cmsq$ 
(consistent with no absorption) and $\Gamma_B=1.37^{+0.79}_{-0.66}$.

The observed range of continuum slopes in radio-quiet quasars is
\hbox{$1.5<\Gamma_{2-10{\rm keV}}<3$}  with a mean of
$\overline{\Gamma}=2.0\pm 0.25$ ($1\sigma$ dispersion; George et 
al. 2000).  Therefore, the very flat slope measured here for
B might be unrealistic, possibly affected by line-of-sight
absorption.  Higher S/N X-ray spectroscopy would reveal whether  
it is the absorption, the intrinsic continua, or both that differ
between the two images.  Only repeated observations, a daunting
investment, can definitively reveal whether the observed differences
are due to variations in a single lensed QSO whose images suffer
differing time delays.  Since the evidence for absorption is weak, if
Q2345+007A,B are indeed lensed, this would predict that a program to
monitor a suitable sample of single luminous QSOs over a period of a
few years would result in the discovery of large temporal variations
in the X-ray continuum slopes of individual
objects.\footnote{Analogous predictions for optical spectroscopic 
variability by SS91 were later confirmed by Small et al. (1997).}
Sufficiently large changes in either the column or continuum slope in
single QSOs have not been observed to date (e.g., Lawson \& Turner
1997).

\subsection{Intrinsic or Intervening Absorbers}
\label{qsoabs}

Could a lens + absorption interpretation explain other data?
From our MOSAIC imaging, or from the colors published by Pello et
al. (1996), the difference between $g^\prime-r^\prime$ and 
$r^\prime-i^\prime$ for the two images is negligible ($-0.032\pm0.02$
and $0.037\pm0.02$, respectively).  The optical colors show little
evidence for reddening relative to the main stellar locus, and indeed
are quite blue in $g^\prime-r^\prime$, as expected for QSOs
at this redshift (Richards et al. 2001).  Thus any putative absorber is
likely to be warm (ionized).  Warm X-ray absorbers are
usually accompanied by a significant decrease in relative X-ray
strength, as might be in evidence in the larger $\aox=1.7$ of Q2345+007B,
but also by evident absorption in the rest-frame ultraviolet spectrum,
particularly in the blue wing of CIV (Brandt, Laor \& Wills 2000). 

High S/N restframe UV spectra of Q2345+007 A and B (SS91) show little
difference between the CIV profiles, and no strong intrinsic
absorption.  Indeed, absorption lines of intervening CIV systems at
$z_{abs}=1.798, 1.799$ and 1.983 are all stronger in A than in B.
Near the CIII] emission line, B shows stronger absorption lines in the
blue wing, while A shows an overall lower profile.  However, this
absorption is most likely due to intervening (not intrinsic)
absorption at $z=1.7717$ (SS91).  Intervening absorbers such as these
in galactic halos or disks have UV columns and ionization levels that
are far too small to cause detectable absorption of the X-ray emission
(O'Flaherty \& Jacobsen 1997).  Interestingly, an intervening CIV
absorber at $z\sim 0.75$ (same as the putative cluster {\bf C1}) would
not show strong lines within the wavelength range of the existing
spectra.  SS91 find evidence for a weak ($\ew\sim1$) Mg\,II absorption
doublet in Q2345+007B at this redshift.  The problem is that an
ionized absorber at $z=0.75$ would be optically bright ($R<22.5$;
Cohen et al. 2000), and easily detected in the deep optical imaging.

The limits on a galaxy that might cause such intervening absorption
are quite stringent ($B\geq 25$ and $H>22.5$; Fischer et al. 1994;
Munoz et al. 1998).  Mg\,II absorbers are usually associated with
galaxies of Hubble types E through Sb (Steidel 1993). Even a normal
$L^*$ Sc galaxy would be 2\,mag brighter than the $H$ magnitude limit 
at this redshift.  An early-type $L^*$ galaxy would be at least 4\,mag
brighter (see the closely related discussion for Q1634+267A,B in Peng et
al. 1999).  So no reasonable candidates exist for a line-of-sight
absorber to account for the X-ray properties of Q2345+007B.

\subsection{Optical/X-ray Flux Ratios of the WSQP}
\label{aox}

For comparison to the optical and infrared bands, the epoch 1990 flux
ratios in $B, R, I, J$ and $K$ bands from Pello et al. (1996) are 3.1,
3.0, 2.9, 4.2, and 4.0 respectively.  The difference between
these flux ratios and those in the X-ray band (6.55, corresponding
to 2.0 on a magnitude scale)  seems significant enough to rule out the
lens interpretation.  

Weir \& Djorgovski (1991) found evidence for variation in the
magnitude difference between the two components, and Gopal-Krishna et
al. (1993) argued that detection of a time delay is crucial to proving
that the WSQP is indeed a lens. If the pair is lensed, then the
expected time delay is about a year.  A comparison with optical flux
ratios more nearly contemporaneous to the X-ray observation is
desired.  The dates of our own optical imaging do not match the
Chandra observations, but begin less than 2 months after.
We compile in  Table~2 the magnitude differences between
Q2345+007 A and B in several optical filters.  Flux ratios extracted
from our (epoch 2000.75) MOSAIC image using SExtractor (Bertin \&
Arnouts 1996) in Sloan $g^\prime$, $r^\prime$, and $i^\prime$ filters
are 5.6, 5.5, and 5.3, respectively.  With errors at most $\sim0.1$
for each ratio, these all agree with each other and suggest no strong
differential reddening of a lensed pair.  However, these optical
ratios still differ from the broadband X-ray flux ratio at the $\sim
2\sigma$ level. Our epoch 2001.63 images show ratios of 4.5, 4.6, 4.4, which
differ at $> 3\sigma$ from the X-ray flux ratio.\footnote{The mean
change in magnitude difference for stars we find in the same filter at
the 2 different epochs is $0\pm0.12$mag.}  The X-ray (0.3-8keV) flux
ratios we measure during the separate  Chandra observations of May and
June 2000 are $8.2\pm 2.1$ and $6.2\pm 1.0$, respectively.  
Given the evidence of variability, we can not rule out the possibility
that at the epoch of the X-ray observations, the optical ratios were
identical to the X-ray flux ratio.

KFM99 developed a statistical proof for comparing the optical and
radio properties of the WSQP population to demonstrate that most WSQPs
must be binary quasars rather than lenses, and that their incidence
was explained by the triggering of quasar activity during galaxy
mergers.  KFM99 classified the quasar pairs as both radio-faint
(denoted $O^2$), both radio-bright ($O^2R^2$), or radio-mixed
($O^2R$). The radio-mixed pairs have wildly discrepant optical/radio
flux ratios (by factors of 50 or more) and are clearly {\em binary}
quasars rather than lenses.  The ratio of radio/optical power spans
$10^4$ in optically-selected quasars, and the radio-loud fraction is
$P_R \sim 10\%$ (e.g., Hooper et al. 1995).  Hence the existence of
each $O^2R$ pair implies the existence $1/2P_R \sim 5$ $O^2$ pairs,
and KFM99 could show statistically that at most 8\% (22\%) of the WSQP
pairs could be gravitational lenses at a 1--$\sigma$ (2--$\sigma$)
limit. These limits still admit a population of optically dark
clusters equal to the normal cluster population, so the possibility of
optically dark clusters still remains.

By imaging an $O^2$ quasar pair with Chandra we can see
if the X-ray and optical flux ratios are discrepant -- an ``$O^2X$''
pair is not a lens.  The only disadvantage of using quasar X-ray emission
is that it is more common than radio emission, and is even 
considered {\em the} signature of an active nucleus.  Nonetheless,  
X-ray to optical flux ratios span a range of $\sim400$ (Pickering,
Impey \& Foltz 1994), so it is probable that binary quasars will show 
discrepancies in their \aox\, values.

Assuming a power-law continuum slope of $\Gamma=2.4$ from spectral
fitting of A and the Milky Way gas column, the observed count rate
corresponds\footnote{We use the Chandra Portable, Interactive
Multi-Mission Simulator (PIMMS; originally Mukai 1993).} to an
(absorbed) X-ray flux in this band for QSO A of $F_{X,A}= 2.52\times
10^{-14}\fcgs
$, while for B the flux is 0.386 in the same units.  From
the unabsorbed values (20\% larger for this slope and $N_H$), the
inferred monochromatic rest-frame luminosities at 2\,keV are
log$L_{2\rm keV}=26.84$ and 26.02 (in erg~s$^{-1}$~Hz$^{-1}$), or
45.11 and 44.30 broadband in \lcgs.

Using the optical $B$ magnitudes from Pello et al. (1996), the
logarithm of the monochromatic 2500\AA\, optical luminosities (in
units \mlcgs) are 31 and 30.5, for A and B respectively. These
correspond to $M_B^A=-26.08$ and $M_B^B=-24.84$ in absolute 
magnitudes at 4400\AA~ (Schmidt \& Green 1983).  We assume an 
optical spectral index of $\alpha=-0.5$, consistent with that found in
SDSS QSOs (with a spread 
of 0.65; Richards et al. 2001), with specific optical normalization
from Marshall et~al. (1984). The resulting \aox\, values
\footnote{\aox\, is the slope of a hypothetical power-law from 
2500\,\AA\, to 2~keV; $\aox\, = 0.384~{\rm log} (\frac{ \lopt}{L_{2\rm
keV} })$.  Use of the nominal best-fit slope of
$\Gamma_B=0.9$ for B does not significantly affect the \aox\, values
($<1\%$ change).} are 1.60 and 1.72.  The value for A is quite normal for
optically-selected QSOs (e.g., Green et al. 1995, Yuan et al. 1998),
while the value of 1.72 for B is more similar to
the values found for absorbed QSOs (Green et al. 2001; Brandt,
Laor \& Wills 2000).  If there is strong absorption along the
sightline to QSO B, it is not detectable in an X-ray spectrum
containing only 55 counts, but more importantly not evident in its
optical colors or restframe UV spectroscopy, as discussed above.

\subsection{Optical Spectra}
\label{ospec}

The exceptionally detailed intrinsic similarity in the optical spectra
of the quasar images (SS91) remains the obstinate core of a
long-running mystery. The differences seen are primarily in 
intervening absorbers, and in the pairs' CIII] emission line
strength.  The intervening absorbers are expected to differ in either
the lens or binary scenario, but more so in the latter. But accounting
for differences in the {\em emission line} profiles is more
problematic. Microlensing by stars in  
the lens could account for $\sim10\%$ differences at most (Schneider
\& Wambsganss 1990).\footnote{The modeling of microlensing effects to date
has assumed Keplerian cloud orbits in the BLR, which yield 
greater predicted differences than infall models.  More recent
outflow disk wind models (Murray \& Chiang 1997; deKool \& Begelman
1995; Elvis 2000) are highly aspect dependent and may yield larger 
differences in microlensing models.}  However, in a lens scenario, the
differences are most likely due to intrinsic profile variations in a
single quasar that occur on a characteristic timescale less than the
time delay between the two images, roughly about a year
(9 months for $z_{lens}=1.5$; $\sim1.8$ years for $z_{lens}=0.75$).
Small et al (1997) confirmed that the observed differences between
emission lines in the Q2345+007 images are consistent with the
spectral differences seen on 1-1.5 year timescales in {\em individual}
QSOs.  So the few observed spectral differences do not rule out
the lens interpretation.  

Given that the current study refutes the lens hypothesis,
is it likely for two {\em distinct} QSOs to have spectra that are as
similar as observed?  Quasar broad emission line strengths and
profiles seem to be loosely correlated with other QSO properties, like
luminosity, $\aox$, and narrow emission line 
strength (e.g., Baldwin 1977; Laor et al. 1997;  Green et al. 2001).
The great similarity in emission lines in Q2345+007 A and B
argues that, if they are distinct objects, their luminosity, the
geometry of their emission regions {\em 
and} their inclination angle should all be similar. Detailed empirical
tests on the probability of achieving such similar spectra at random
among a sample of unrelated 
quasars require large databases of quasar spectra, uniformaly
analyzed, and with similar S/N and resolution.  Such studies
are beyond the scope of this paper, but adequate measurement datasets
are now becoming available (Forster et al. 2000).

\section{Summary}
\label{summary}

We conclude that Q2345$+$007A,B is a quasar binary.  

First, we find {\em no} evidence for a halo in the field sufficiently
massive to produce the image separation, provided that halo contains
at least 1/3 of the normal cluster baryon fraction.  Mechanisms
may exist to suppress star formation in galaxies (e.g.  Elmegreen \&
Parravano 1994) or to produce only exceptionally low surface
brightness galaxies (e.g. Sprayberry et al. 1995) thereby diminishing
the optical luminosity of cluster members.  There is no such mechanism
known that could significantly reduce the baryon fraction in a cluster
potential well. The energy injection from star formation or AGN would
need to be adequate not only to heat and eject gas from the galaxies,
but also from the massive cluster potential. Even if the energetics
were plausible, we are unable to find any of the concomitant optical
or infrared emission (e.g. Mcleod, Rieke, \& Weedman 1994). 

Second, the two quasars have X-ray properties inconsistent with the
lens hypothesis.  The X-ray flux ratios are inconsistent
with the optical flux ratios.  Their X-ray spectral properties also
differ significantly, and we find no evidence for either
extinction or absorption that might explain the inconsistency in terms
of the lens hypothesis.

Significantly deeper spatially resolved X-ray spectroscopy could show
more definitively that the pair is a binary.  The best available hope
to prove the pair is lensed would be detection of a time-delay in
correlated variability.  Optical or near-IR detection of host
galaxies in the QSO images could also provide strong arguments for or
against the lens hypothesis.

The X-ray evidence in the field of Q2345+007 does not support
the lens hypothesis.  This WSQP is thus likely to be a binary quasar,
and currently stands as the example with the highest redshift, the
largest separation, and the most detailed agreement between optical/UV
spectra of its components.  We may be studying a pair of luminous QSOs
whose hosts, separated by $\sim 60h_{50}$kpc, have a history of
dynamical interaction.  Plausible models hold that quasar activity 
is triggered by tidal interactions in a galactic merger, but that
activation of the galactic nuclei occurs late in the interaction, when
the nuclei are within $80\pm30$kpc of each other (Mortlock et
al. 1999).  Simple dynamical friction models reproduce the observed
distribution of projected separations of pairs, but predict that
binary quasars are only observable as such in the early stages of
galactic collisions, after which the supermassive black holes would
orbit within the merger remnant. 
We speculate that Q2345+007 may represent the highest redshift
example known of interaction-triggered but as-yet unmerged luminous
AGN.   

\bigskip
We thank Frank Valdes, Lindsey Davis, and the IRAF team for writing
and helping with the MSCRED package. Thanks also to Brian MacLean at
STScI for early help with the GSC2.2 catalog.  
Thanks to Ani Thakar at Johns Hopkins for help retrieving SDSS data.
Funding for the creation and distribution of the SDSS Archive has been
provided by the Alfred P. Sloan Foundation, the Participating
Institutions, the National Aeronautics and Space Administration, the
National Science Foundation, the U.S. Department of Energy, the
Japanese Monbukagakusho, and the Max Planck Society. The SDSS Web site
is http://www.sdss.org/.

This work was supported
by CXO grant GO~0-1161X and NASA grant NAS8-39073. AD, PJG, DK, MM, and AS
acknowledge support through NASA  Contract NASA contract NAS8-39073
(CXC).  CSK is supported by NASA grants NAG5-8831 and NAG5-9265.

\begin{figure*}[ht]
\figurenum{4}
\plotone{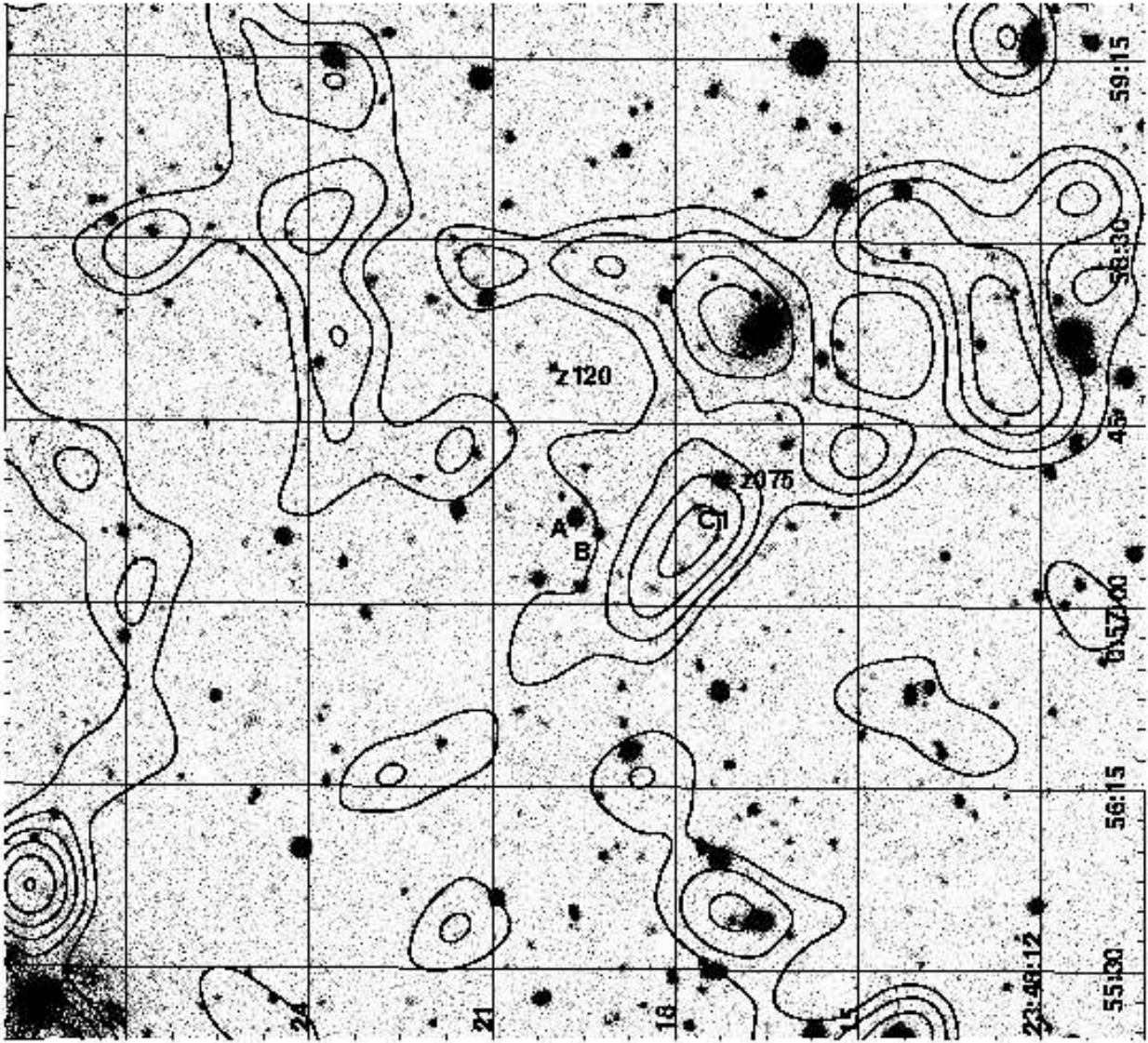}
\caption{An $r^{\prime}$ image from the CTIO 4meter from
UT 29 September 2000 of a $4.25\arcmin$ field surrounding Q2345$+$007.
North is up, East to the left, and a J2000 coordinate grid
displayed.  The large contours show the same X-ray flux levels as in
Figure~\ref{fnosrc}.  Positions of the QSO images A and B
are and of putative optically-identified galaxy clusters are  marked
in bold type.  
\label{fnosrcopt} }
\end{figure*}

\begin{figure*}[ht]
\figurenum{5}
\centering
\vspace*{0.0in}
\plotone{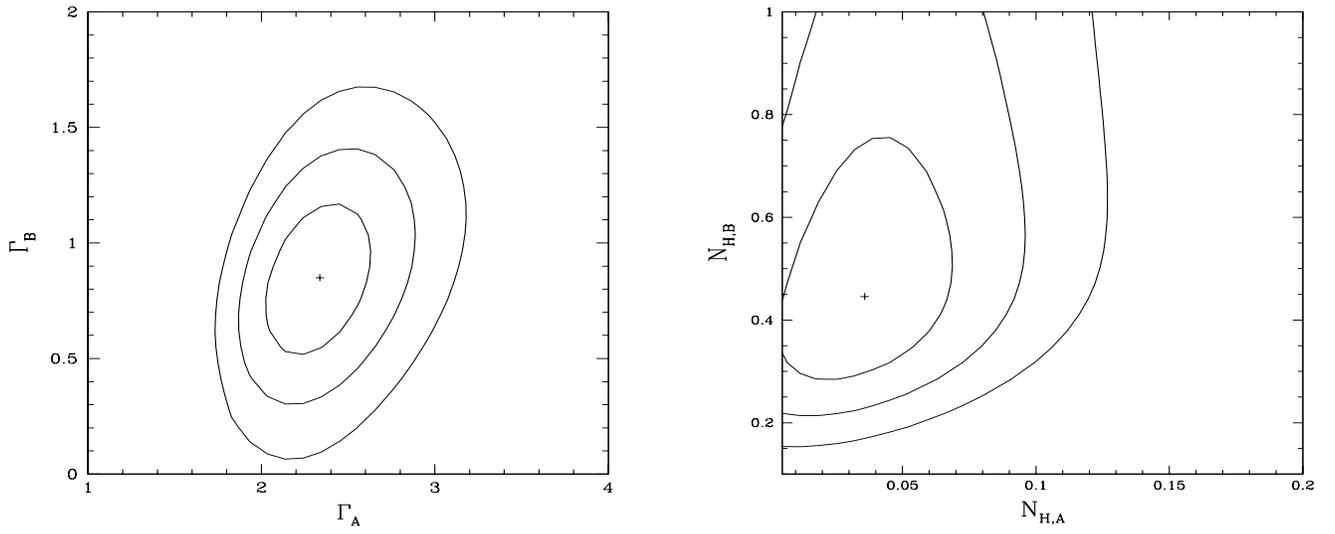}
\caption{\small 
Contour plots of confidence levels for simultaneous fits to ACIS
spectra of Q2345+007 A and B.  The plot at left shows the confidence
levels for $\Gamma_A$ vs. $\Gamma_B$ in Model (2) where
the absorption is assumed to be common.  The plot at right shows
confidence levels for $N_{H,A}$ vs. $N_{H,B}$ in Model (3) where the
power-law slope is assumed to be common.   
\label{fspeccont}}
\end{figure*}

\clearpage

\begin{deluxetable}{lllcc}
\tablefontsize{\footnotesize}
\tablewidth{300pt}
\tablenum{1}
\tablecaption{Measured Positions of QSOs and Possible Cluster Centers}
\label{tcenters}
\tablehead{ Center  &  \multicolumn{2}{c}{RA (2000) Dec\tablenotemark{a}}  &
Separation\tablenotemark{b} &  Reference } 
\startdata
        & \multicolumn{2}{c}{Optical}&  &  \\ 
QSO A,O &  23 48 19.6 &  00 57 21.6  & ... & 1 \\
QSO B,O &  23 48 19.2 &  00 57 17.7  & ... & 1 \\
center  &  23 48 19.4 &  00 57 19.2  & ... &  1 \\
        &             &              &     &    \\
C1      &  23 48 17.2 &  00 57 22    & 33  &  2 \\
G1      &  23 48 16.5 &  00 58 08.3  & 66  &  2 \\
G2      &  23 48 17.2 &  00 57 31.1  & 36  &  2 \\
Z075    &  23 48 16.6 &  00 57 31    & 44  &  3 \\ 
Z120    &  23 48 19.3 &  00 57 57    & 38  &  3 \\
        &             &              &     &    \\
        & \multicolumn{2}{c}{X-ray}  &     &  \\ 
QSO A   &  23 48 19.6 &  00 57 21.4  & 0.3 &  1 \\
QSO B   &  23 48 19.2 &  00 57 17.5  & 0.2 &  1 \\
CXO J234817.6+005717
        &  23 48 17.6 &  00 57 17  & 27 &  1 \\
CXO J234816.9+005811
        &  23 48 16.9 &  00 58 11  & 64 &  1 \\
CXO J234812.7+005813
       &   23 48 12.7 &  00 58 13  & 114 &  1 \\
\tablenotetext{a}{Coordinates with subarcsecond precision
are listed for objects with well-defined centroids from our optical
imaging.}  
\tablenotetext{b}{For QSOs, separation of X-ray centroids from optical
centroids. For putative optical clusters, separations are in arcsec
from the center of the QSO pair center.} 
\tablerefs{(1) This paper; (2) Bonnet et al. 1993;
(3) Pello et al. 1996. }  
\enddata
\end{deluxetable}

\begin{deluxetable}{cclccclc}
\tablefontsize{\footnotesize}
\tablewidth{340pt}
\tablenum{2}
\tablecaption{Magnitude differences for Q2345+007 A and B}
\label{tratios}
\tablehead{ Epoch &  Band  &  $m_B-m_A$  & Reference &  Epoch &  Band  &  $m_B-m_A$  & Reference  }
\startdata
1981.92  & $B$  & 1.43$\pm0.06$  &  5 & 1991.96  & $R$   & 1.25$\pm0.09$    &  4   \\      
         & $V$  & 1.43$\pm0.05$  &  5 &	1992.73  & $V$   & 1.13$\pm0.12$    &  4   \\      
1981.98  & $r$  & 1.54$\pm0.13$  &  5 &	1992.88  & $B$   & 1.30$\pm0.06$    &  4   \\      
1982.57  & $r$  & 1.44$\pm0.15$  &  5 &	1992.93  & $K$   & 1.3$\pm0.1$   &  7   \\         
1982.90  & $r$  & 1.41$\pm0.13$  &  5 &	1993.76  & $J$   & 1.55    &  3   \\               
1989.65  & $B$  & 1.30$\pm0.14$  &  6 &	1993.76  & $K^{\prime}$ & 1.51 &  3   \\           
         & $g$  & 1.31$\pm0.07$  &  6 &	1998.73  & $u^{\ast}$  & 1.62$\pm 0.12$  &  8   \\ 
         & $r$  & 1.24$\pm0.06$  &  6 &	         & $g^{\ast}$  & 1.63$\pm 0.03$  &  8   \\ 
         & $i$  & 1.08$\pm0.06$  &  6 &	         & $r^{\ast}$  & 1.60$\pm 0.05$  &  8   \\ 
1989.67  & $r$  & 1.28$\pm0.04$  &  6 &	         & $i^{\ast}$  & 1.62$\pm 0.07$  &  8   \\ 
1989.69  & $g$  & 1.32$\pm0.03$  &  6 &	         & $z^{\ast}$  & 1.37$\pm 0.16$  &  8   \\ 
         & $r$  & 1.27$\pm0.04$  &  6 &	2000.75  & $g^{\ast}$  & 1.88$\pm0.01$  &  1   \\  
         & $i$  & 1.19$\pm0.04$  &  6 &	         & $r^{\ast}$  & 1.85$\pm0.01$  &  1   \\  
1989.96  & $K$  &  1.11$\pm0.17$ &  4 &	         & $i^{\ast}$  & 1.81$\pm0.01$  &  1   \\  
1990.79  & $B_J$ & 1.24          &  3 &	2001.63  & $g^{\ast}$  & 1.64$\pm0.01$  &  1   \\  
1990.79  & $R$   & 1.19          &  3 &	         & $r^{\ast}$  & 1.65$\pm0.01$  &  1   \\  
1990.79  & $I$   & 1.14          &  3 &	         & $i^{\ast}$  & 1.60$\pm0.01$  &  1   \\  

\tablerefs{(1) This paper; (2) Bonnet et al. 1993;
(3) Pello et al. 1996; (4) Gopal-Krishna et al. 1993; (5) Sol et
al. 1984; (6) Weir \& Djorgovski 1991; (7) McLeod et al. 1994;
(8) SDSS Early Data Release.}  
\tablecomments{We include only ratios published with accurate dates
in {\em B} band or redder.} 
\enddata
\end{deluxetable}

\begin{deluxetable}{lclll}
\tablefontsize{\footnotesize}
\tablewidth{240pt}
\tablenum{3}
\tablecaption{Spectral Fit Parameters}
\label{tfit}
\tablehead{
 Model & QSO  &  $\Gamma$  &  $N_H$   &  $\chi^2$ (DOF)\tablenotemark{a} \\
       &      &            &  ($10^{20}$~cm$^{-2}$) &  \\}
\startdata
1 & A & $2.19\pm0.15$ & 3.8  & 42.54 (40) \vspace{0.5ex} \\
  & B & $0.79\pm0.4$  & ...  & 25.09 (40) \vspace{0.5ex} \\
2 & A & $2.30^{+0.36}_{-0.30}$ & $5.3\pm3.1$  & 67.9 (79) \vspace{0.5ex} \\
  & B & $0.83^{+0.49}_{-0.44}$ & ...  & ...       \vspace{0.5ex} \\
3 & A & $2.14^{+0.34}_{-0.28}$ & $3.4^{+3.5}_{-2.9}$ & 65.4 (79) \vspace{0.5ex} \\
  & B & ...  & $43.9^{+29}_{-18}$ & ...       \vspace{0.5ex} \\
4 & A & $2.23^{+0.33}_{-0.30}$ & $4.3\pm3.1$  & 42.77 (39) \vspace{0.5ex} \\
  & B & $1.37^{+0.79}_{-0.66}$ & $23.6^{+28.1}_{-19.2}$  & 21.32 (39)\vspace{0.5ex} \\
\tablecomments{Fit parameters based on simultaneous fitting of
spectra using Primini statistics in Sherpa (Freeman et al 2001).
Uncertainties are 90\% confidence limits.  Where no uncertainties are
shown, the parameter was frozen at the value displayed.  Where only
QSO A shows a value, the parameter was fit simultaneously to both
components A and B.  Models are described in the text.}
\tablenotetext{a}{$\chi^2$ based on spectra
binned to 10 counts per bin, using given fit parameters.}
\enddata
\end{deluxetable}

\end{document}